\documentclass[aps,prd,reprint,nofootinbib,superscriptaddress,floatfix,longbibliography]{revtex4-2}
\usepackage{amsmath,amssymb,bm}
\usepackage{graphicx,booktabs}
\usepackage{xcolor}
\usepackage[colorlinks=true,linkcolor=blue,citecolor=blue,urlcolor=blue]{hyperref}
\graphicspath{{figures/}}
\newcommand{\dd}{\mathrm{d}}
\newcommand{\KL}{D_{\mathrm{KL}}}
\newcommand{\E}{\mathbb{E}}
\newcommand{\rhob}{\bm{\rho}}
\newcommand{\HN}{\operatorname{HalfNormal}}
\hypersetup{pdftitle={An Information-Theoretic Perspective on Spectral Reconstruction in Lattice QCD}}

\begin{document}
\title{An Information-Theoretic Perspective on Spectral Reconstruction in Lattice QCD}

\author{Nan Wang} %
\affiliation{State Key Laboratory of Nuclear Physics and Technology, Institute of Quantum Matter, South China Normal University, Guangzhou 510006, China}
\affiliation{Key Laboratory of Atomic and Subatomic Structure and Quantum Control~(MOE), Guangdong-Hong Kong Joint Laboratory of Quantum Matter, Guangzhou 510006, China}
\affiliation{Guangdong Basic Research Center of Excellence for Structure and Fundamental Interactions of Matter, Guangdong Provincial Key Laboratory of Nuclear Science, Guangzhou 510006, China}
\author{Long-cheng Gui} %
\email{guilongcheng@hunnu.edu.cn}
\affiliation{Department of Physics, Hunan Normal University,  Changsha 410081, China }
\affiliation{Synergetic Innovation
Center for Quantum Effects and Applications (SICQEA), Changsha 410081, China}
\affiliation{Key Laboratory of Low-Dimensional Quantum Structures and Quantum Control of Ministry of Education, Changsha 410081, China}
\author{Jian Liang} %
\email{jianliang@scnu.edu.cn}

\affiliation{State Key Laboratory of Nuclear Physics and Technology, Institute of Quantum Matter, South China Normal University, Guangzhou 510006, China}
\affiliation{Guangdong Basic Research Center of Excellence for Structure and Fundamental Interactions of Matter, Guangdong Provincial Key Laboratory of Nuclear Science, Guangzhou 510006, China}
\author{Jun Shi} %
\email{jun.shi@scnu.edu.cn}
\affiliation{Guangdong Provincial Key Laboratory of Sensing Physics and System Integration Applications,
School of Physics and Optoelectronic Engineering, Guangdong University of Technology,\\
Guangzhou, Guangdong 510006, China\\~\\
\includegraphics[scale=0.1]{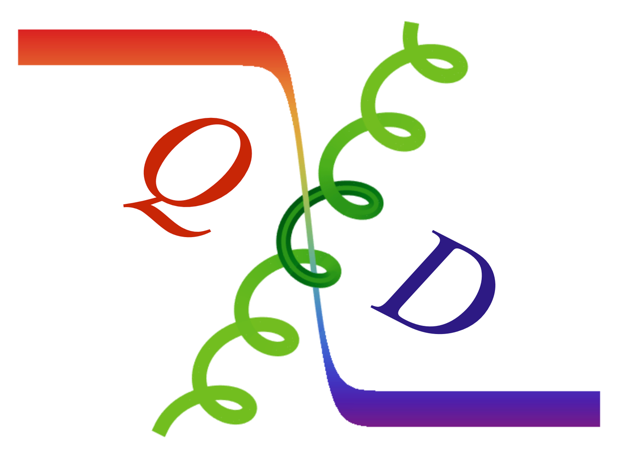}\\
{\rm($\chi$QCD Collaboration)}}
\author{Han-yang Xing} %
\affiliation{State Key Laboratory of Nuclear Physics and Technology, Institute of Quantum Matter, South China Normal University, Guangzhou 510006, China}
\affiliation{Guangdong Basic Research Center of Excellence for Structure and Fundamental Interactions of Matter, Guangdong Provincial Key Laboratory of Nuclear Science, Guangzhou 510006, China}

\begin{abstract}
Assessing the reliability of spectral reconstruction from finite, noisy
lattice correlators requires understanding the constraints supplied by
the data relative to prior assumptions. We develop an
information-theoretic framework using Kullback--Leibler (KL)
divergences to quantify the
information supplied by the data through changes from prior to
posterior probability distributions. Our analysis accounts for both
changes in the probability distributions of spectral strength at
individual energies and constraints encoded in correlations across
the spectrum. We smear these information measures over neighboring
energies using Gaussian weights to obtain a smoothed assessment at a
specified energy scale.
We demonstrate this framework using a hierarchical Bayesian
reconstruction method with posterior sampling, applied to mock correlator
data generated from a three-peak spectrum and a near-physical lattice
proton correlator. These
calculations show that the information measures help explain the
properties of reconstructed spectra. They also provide a semiquantitative
criterion for assessing how strongly reconstructed spectral structures
are constrained by the data and how reliable they are in light of
prior effects.
\end{abstract}
\maketitle

\section{Introduction}
\label{sec:introduction}

Reconstructing spectral functions from Euclidean correlation functions
is a central inverse problem in lattice QCD, providing access to
real-time dynamics and inclusive observables of direct phenomenological
interest. Applications include transport properties and thermal photon
production in hot QCD~\cite{Meyer2011,Ali2024}, the hadronic $R$-ratio
in $e^+e^-$ annihilation~\cite{Alexandrou2023R}, the hadronic tensor
governing lepton-nucleon scattering~\cite{Liang2020,Liang2026}, and
inclusive decays into hadronic final states, such as hadronic $\tau$
decays~\cite{Alexandrou2024Tau} and semileptonic $D_s$ decays~\cite{DeSantis2025PRL,DeSantis2025,Kellermann2026}. Spectral methods also
offer access to long-distance contributions to electroweak amplitudes~\cite{Frezzotti2023,Frezzotti2026}. These applications call for spectral
functions or their weighted integrals that incorporate the contributions
of allowed intermediate or final states without resolving them
individually, going beyond the extraction of the energies and matrix
elements of a few isolated states through exponential fits~\cite{Hashimoto2026}. Treating these contributions collectively also
offers a route to addressing finite-volume effects in multihadron
systems~\cite{Hansen2017}. These considerations underscore the importance
of reliably solving the inverse problem posed by finite, noisy Euclidean
correlator data.

A variety of methods have been applied or developed to address this
inverse problem, including the maximum entropy method (MEM)~\cite{Asakawa2001},
Bayesian reconstruction (BR)~\cite{Burnier2013}, and the Backus-Gilbert
(BG) method~\cite{Backus1968,Hansen2017}. These approaches have been
complemented by the Hansen-Lupo-Tantalo (HLT) method~\cite{Hansen2019} and Chebyshev expansions~\cite{Bailas2020}, as well as
Gaussian-process regression~\cite{Horak2022,DelDebbio2025}, sparse
modeling~\cite{Itou2020}, and neural networks~\cite{Kades2020,Buzzicotti2024,Wang2022}. Further developments include
analytic formulations using conformal maps~\cite{Bergamaschi2023} and
Mellin transforms~\cite{Bruno2025}, together with convex optimization
for bounding spectral integrals~\cite{Abbott2026}. Posterior sampling
has also enabled Bayesian analyses beyond a single optimized spectrum~\cite{Rothkopf2022}. Despite these
methodological advances, a central question remains: how reliably can
a spectrum be reconstructed from a finite set of correlator measurements
with statistical uncertainties? Assessing this reliability requires
understanding the relative roles of the data and of priors or other
assumptions, as well as how the strength of the data constraints varies across
energy regions.

Information theory provides a natural framework for addressing this
question. By quantifying how the data update a prior distribution to
a posterior distribution, it characterizes the additional constraints
supplied by the data relative to the assumptions of the reconstruction.
This analysis can examine both changes in the marginal distributions
at individual energies and joint constraints involving different energies.
These comparisons help identify which parts of the spectrum are
constrained by the data and which remain largely determined by the prior,
thereby informing the assessment of reliability across energy regions.

To implement this perspective, we use the conditional Kullback-Leibler (KL)
divergence~\cite{Kullback1951} to quantify
the marginal update at each energy while accounting for the inferred
hyperparameter distribution. We also introduce a diagnostic
of joint constraints, which is constructed from full conditional
distributions and decomposes into a marginal KL divergence and a posterior
mutual-information contribution. This decomposition separates changes
in individual marginal distributions from constraints encoded in
correlations across the spectrum. We evaluate these measures
within a hierarchical BR model using both mock data and lattice QCD
correlator data, providing a basis for a semiquantitative assessment of
reconstruction reliability.

Section~\ref{sec:information} presents the information-theoretic
framework, and Sec.~\ref{sec:implementation} describes the hierarchical
BR reconstruction and numerical evaluation of $K$ and $J$ from
posterior samples. Section~\ref{sec:mock} examines the framework using
mock data generated from a known three-peak spectrum.
Section~\ref{sec:real} applies it to a near-physical lattice proton
correlator to assess the data constraints and prior influence across
energy regions. Section~\ref{sec:summary} summarizes the main findings
and outlines future directions.

\section{Information-theoretic framework}
\label{sec:information}

To assess how strongly the data constrain different parts of a reconstructed
spectrum, we compare the posterior density $p$ with the prior density $q$
using the KL divergence,
\begin{equation}
 \KL(p\Vert q)=\int\dd x\,p(x)\log_2\frac{p(x)}{q(x)}.
 \label{eq:kl}
\end{equation}
We use logarithms to base two, so all information quantities are in bits.
This divergence measures the information supplied by the data relative
to the prior. It is asymmetric and has no universal upper bound.

To apply KL to spectral reconstruction, we first
specify the prior and posterior distributions of the spectrum.
Let $D$ denote the correlator data and
$\rhob=(\rho_1,\ldots,\rho_{N_\omega})$ the spectrum evaluated
at discrete energies $\omega_i$, with $\rho_i=\rho(\omega_i)$. We introduce
a shared hyperparameter $\alpha>0$ that controls the spectral prior.
The joint posterior of $\rhob$ and $\alpha$ is
\begin{equation}
 p(\rhob,\alpha\mid D)
 =\frac{p(D\mid\rhob)\,p(\rhob\mid\alpha)\,p(\alpha)}{p(D)}.
 \label{eq:bayes}
\end{equation}
The KL chain rule separates the information supplied by the data into
an update of $\alpha$ and a spectral update at fixed $\alpha$:
\begin{align}
 &\KL\bigl[p(\rhob,\alpha\mid D)\Vert p(\rhob,\alpha)\bigr]
 \nonumber\\
 &=\KL\bigl[p(\alpha\mid D)\Vert p(\alpha)\bigr]
 \nonumber\\
 &\quad+\E_{\alpha\mid D}\Bigl[\KL\bigl[
 p(\rhob\mid D,\alpha)\Vert p(\rhob\mid\alpha)\bigr]\Bigr],
 \label{eq:hierarchy}
\end{align}
where $\E_{\alpha\mid D}$ denotes averaging over the posterior
$p(\alpha\mid D)$, namely
\[
 \E_{\alpha\mid D}[f(\alpha)]
 =\int_0^\infty\dd\alpha\,p(\alpha\mid D)f(\alpha).
\]
We focus on the spectral update in the second term of
Eq.~\eqref{eq:hierarchy}: the change from prior to posterior at fixed
$\alpha$, averaged over $p(\alpha\mid D)$. To examine the constraints
at each energy $\omega_i$, we integrate out the spectral values at all
other energies to obtain the marginal distribution of $\rho_i$.
We then define the conditional KL as
\begin{equation}
 K_i\equiv K(\omega_i)=\E_{\alpha\mid D}\Bigl[\KL\bigl[
 p(\rho_i\mid D,\alpha)\Vert p(\rho_i\mid\alpha)\bigr]\Bigr].
 \label{eq:ki}
\end{equation}
The energy-dependent measure $K$, with values $K_i$ at $\omega_i$,
measures the information supplied by the data about the spectral value
at each energy relative to its conditional prior, as expressed in the
marginal distribution.
Conditioning on $\alpha$ avoids counting a change caused solely by its update
as a separate constraint at every energy.

Correlator data constrain weighted integrals of the spectrum and thus
couple spectral values at different energies. These correlations are
not captured by the marginal comparisons above. To include them, we
introduce a full-conditional KL diagnostic $J$, with values
\begin{equation}
 \begin{aligned}
 J_i\equiv J(\omega_i)
 &=\E_{\alpha,\rhob_{-i}\mid D}\Bigl[\\
 &\quad\KL\bigl[p(\rho_i\mid D,\alpha,\rhob_{-i})
 \Vert p(\rho_i\mid\alpha)\bigr]\Bigr],
 \end{aligned}
 \label{eq:ji}
\end{equation}
where $\rhob_{-i}=(\rho_1,\ldots,\rho_{i-1},\rho_{i+1},\ldots,\rho_{N_\omega})$
is the vector of spectral values with $\rho_i$ omitted. The full
conditional describes the allowed variation of $\rho_i$ for given
$\alpha$ and $\rhob_{-i}$, while the expectation averages
over their joint posterior to account for their uncertainty.
Adding and subtracting $\log_2 p(\rho_i\mid D,\alpha)$ inside this
expectation relates $J_i$ to the marginal KL:
\begin{equation}
 J_i=K_i+\E_{\alpha\mid D}\Bigl[
 I_{p(\rhob\mid D,\alpha)}(\rho_i;\rhob_{-i})\Bigr].
 \label{eq:decomposition}
\end{equation}
Here $I$ is the mutual information between $\rho_i$ and the remaining
spectrum in the posterior at fixed $\alpha$. For a joint density
$p(x,y)$, it is defined as
\begin{equation}
 I_p(X;Y)=\KL\bigl[p(x,y)\Vert p(x)p(y)\bigr].
 \label{eq:mutual_information_definition}
\end{equation}
Mutual information quantifies the average reduction in posterior
uncertainty about $\rho_i$ when the spectral values at all other energies
are known, at fixed $\alpha$.
Its nonnegativity implies $J_i\geq K_i$.
Thus, in our reconstruction, $J_i$ measures the information supplied by
the data about $\rho_i$ relative to its conditional prior, including both
the marginal information $K_i$ and information encoded in correlations
with the rest of the spectrum.

The additional term in Eq.~\eqref{eq:decomposition} has a direct physical
interpretation. At fixed $\alpha$, our prior factorizes, so posterior
correlations between spectral values arise from the data constraints.
For example, an increase in spectral weight at one energy may require
compensating changes at other energies to remain compatible with the
correlator data. A small $K_i$ with an appreciable $J_i$ then indicates
that the data restrict the allowed combinations of spectral values even
though the marginal distribution of $\rho_i$ remains close to its prior.

\section{Bayesian reconstruction and numerical evaluation}
\label{sec:implementation}

We use a PyMC implementation of the BR method~\cite{Burnier2013} as an
example to illustrate and test our information-theoretic approach.
This section describes the reconstruction setup, posterior sampling and the detailed
methods for computing the information measures used in the mock and
lattice data studies presented in the following sections.

\subsection{Forward model and spectral prior}

The spectral representation of the Euclidean correlator and its discrete
form are
\begin{equation}
 C(t)=\int_0^\infty\dd\omega\,e^{-\omega t}\rho(\omega),
 \qquad C_t=\sum_i K_{ti}\rho_i,
 \label{eq:forward}
\end{equation}
where the subscript $t$ labels discrete Euclidean times and $i$ labels
the discrete energies $\omega_i$. We approximate the integral over a
finite energy interval using $K_{ti}=\Delta\omega_i e^{-\omega_i t}$,
with $\Delta\omega_i$ the integration weight associated with $\omega_i$.
For the measured ensemble means $\bar C_t$, we use a Gaussian
likelihood,
\begin{equation}
 p(D\mid\rhob)\propto
 \exp\left[-\tfrac12(\bar C-K\rhob)^T\Sigma^{-1}(\bar C-K\rhob)\right].
 \label{eq:likelihood}
\end{equation}
Here $\Sigma$ is the covariance matrix of the ensemble means, including
correlations between Euclidean times.

For the spectral prior, we use the BR functional~\cite{Burnier2013},
\begin{equation}
 T(\rhob)=\sum_i \Delta\omega_i\left[\frac{\rho_i}{m_i}-1-
 \ln\!\left(\frac{\rho_i}{m_i}\right)\right],
 \label{eq:br_functional}
\end{equation}
where $m_i=m(\omega_i)>0$ is the default model.
The corresponding prior is
$p(\rhob\mid\alpha)\propto\exp[-\alpha T(\rhob)]$, with the positive
hyperparameter $\alpha$ controlling its strength relative to the
likelihood. Increasing $\alpha$ strengthens the penalty for deviations
from the default model.
Since $T$ is a sum over individual energies, this prior factorizes at
fixed $\alpha$. Normalizing each factor gives
\begin{align}
 p(\rhob\mid\alpha)&=\prod_i p(\rho_i\mid\alpha),
 \nonumber\\
 p(\rho_i\mid\alpha)&=
 \frac{(c_i/m_i)^{1+c_i}}{\Gamma(1+c_i)}
 \rho_i^{c_i}e^{-c_i\rho_i/m_i},\qquad c_i=\alpha\Delta\omega_i.
 \label{eq:gamma}
\end{align}
Each spectral value $\rho_i>0$ thus follows a Gamma distribution with
shape $1+c_i$ and rate $c_i/m_i$.
The conditional prior has mode and mean
\begin{equation}
 \rho_{i,\mathrm{mode}}=m_i,\qquad
 \E[\rho_i\mid\alpha]=m_i\left(1+\frac{1}{\alpha\Delta\omega_i}\right).
 \label{eq:prior_moments}
\end{equation}

We assign $\alpha$ a HalfNormal prior with scale $\sigma_\alpha$,
$\alpha\sim\HN(\sigma_\alpha)$, obtained by restricting a zero-mean
Gaussian distribution to $\alpha>0$ and renormalizing. Its density is
\begin{equation}
 p(\alpha)=\sqrt{\frac{2}{\pi}}\,\frac{1}{\sigma_\alpha}
 \exp\left[-\frac{\alpha^2}{2\sigma_\alpha^2}\right],
 \qquad \alpha>0.
 \label{eq:hyperprior}
\end{equation}
The default model $m_i$, energy discretization $\Delta\omega_i$ and
hyperprior scale $\sigma_\alpha$ are set separately for each data set
and reconstruction setup.

\subsection{Posterior sampling and information measures}

With the likelihood and priors specified above, we sample the joint
posterior $p(\rhob,\alpha\mid D)$ in PyMC using the NumPyro No-U-Turn
sampler~\cite{Hoffman2014}. Each production run uses ten chains, with
1000 warm-up steps and 1000 retained draws per chain. These joint samples
provide the posterior distributions of both $\rhob$ and $\alpha$ while
preserving their correlations.

The posterior distribution allows us to characterize the reconstructed
spectrum through its mean, credible intervals and correlations.
Alongside the posterior mean, we consider the whole-spectrum maximum
a posteriori (MAP) estimate,
\begin{equation}
 \rhob_{\rm MAP}=\underset{\rhob>0}{\operatorname{arg\,max}}
 p(\rhob\mid D),
 \label{eq:map}
\end{equation}
where $p(\rhob\mid D)$ is the spectral posterior with $\alpha$
marginalized out. We compare the posterior mean and MAP spectra in the
numerical results.

According to Eqs.~\eqref{eq:ki} and~\eqref{eq:ji}, evaluating $K$ and
$J$ requires the marginal posterior $p(\rho_i\mid D,\alpha)$ and the
full-conditional posterior $p(\rho_i\mid D,\alpha,\rhob_{-i})$,
respectively, together with the conditional prior specified above.
The full-conditional posterior follows directly from the Gaussian
likelihood and BR prior:
\begin{equation}
 p(\rho_i\mid D,\alpha,\rhob_{-i})
 \propto \rho_i^{c_i}\exp\left[-\tfrac12a_i\rho_i^2+
 (b_i-c_i/m_i)\rho_i\right],
 \label{eq:fullconditional}
\end{equation}
where the coefficients are
\begin{equation}
 \begin{aligned}
 a_i&=\sum_{t,t'}K_{ti}(\Sigma^{-1})_{tt'}K_{t'i},\\
 b_i&=\sum_{t,t'}K_{ti}(\Sigma^{-1})_{tt'}
 \left(\bar C_{t'}-\sum_{j\ne i}K_{t'j}\rho_j\right).
 \end{aligned}
 \label{eq:conditional_coeff}
\end{equation}
For given $\alpha$ and $\rhob_{-i}$, $\rho_i$ remains a free variable.
We normalize this density over $\rho_i>0$ by one-dimensional integration.

The marginal posterior required for $K$ is obtained by integrating
over the remaining spectral values:
\begin{equation}
 \begin{aligned}
 p(\rho_i\mid D,\alpha)
 &=\int\dd\rhob_{-i}\,p(\rho_i\mid D,\alpha,\rhob_{-i})\\
 &\qquad\times p(\rhob_{-i}\mid D,\alpha).
 \end{aligned}
 \label{eq:marginal_from_conditionals}
\end{equation}
Each joint sample contains one complete spectrum $\rhob$ and one value
of $\alpha$. We sort these
paired samples by $\alpha$ and divide the ordered sequence into 20 bins
with equal sample counts. Within each bin, the sampled values of
$\rhob_{-i}$ represent the probability weight
$p(\rhob_{-i}\mid D,\alpha)$ in
Eq.~\eqref{eq:marginal_from_conditionals}, so the integral is
approximated by the arithmetic mean
\begin{equation}
 p(\rho_i\mid D,\alpha)\simeq\frac{1}{n_B}\sum_{s\in B}
 p(\rho_i\mid D,\alpha^{(s)},\rhob_{-i}^{(s)}),
 \label{eq:conditional_sample_average}
\end{equation}
where $B$ denotes an $\alpha$ bin containing $n_B$ joint posterior
samples, labeled by $s$.
Each sample supplies $\alpha^{(s)}$ and $\rhob_{-i}^{(s)}$; the density
in each term is calculated from Eq.~\eqref{eq:fullconditional} as a
function of $\rho_i$, rather than estimated from that sample's value
of $\rho_i$.
The average includes the small variation of $\alpha$ within the bin and
approximates the marginal density at fixed $\alpha$.

For the prior in the same bin, we average the Gamma densities in
Eq.~\eqref{eq:gamma} over the bin's $\alpha$ samples. We denote the KL
divergence between the resulting posterior and prior densities at
energy $\omega_i$ in bin $B$ by $K_{B,i}$. Following the posterior
average over $\alpha$ in Eq.~\eqref{eq:ki}, we obtain
\begin{equation}
 \begin{aligned}
 K_i&\simeq\sum_B
 \left[\int_{\alpha\in B}\dd\alpha\,p(\alpha\mid D)\right]K_{B,i}\\
 &\simeq\frac{1}{20}\sum_B K_{B,i},
 \end{aligned}
 \label{eq:ki_bin_average}
\end{equation}
where $1/20$ is the estimated posterior probability for $\alpha$ to lie
in each bin, since each bin contains 500 of the 10000 posterior samples
sorted by $\alpha$.

Similarly, $J_i$ follows from Eq.~\eqref{eq:ji} by averaging the KL
divergence of each normalized conditional density in
Eq.~\eqref{eq:fullconditional} relative to its corresponding prior
$p(\rho_i\mid\alpha)$ over the joint posterior samples.

We check the normalization of the averaged posterior densities in
Eq.~\eqref{eq:conditional_sample_average} and verify the convergence of
the numerical integrals by increasing the integration resolution.

\section{Mock-data results}
\label{sec:mock}

\subsection{Setup and reconstructed spectra}

Our mock spectrum contains three equal-weight peaks,
\begin{equation}
 \rho_{\rm true}(\omega)=\sum_{n=1}^3\delta(\omega-E_n),
 \qquad (E_1,E_2,E_3)=(1,1.5,3).
 \label{eq:mock_truth}
\end{equation}
The corresponding exact correlator is $C(t)=\sum_n e^{-E_n t}$.
We generate 81 correlated Gaussian samples, with the relative statistical
error of their mean increasing from approximately $0.07\%$ at $t=1$ to
$0.54\%$ at $t=15$, yielding a high-precision correlator.
We reconstruct the spectrum from the ensemble mean
and its estimated full covariance matrix over $t=1,\ldots,15$.
For the baseline analysis, we use
$m_i=m=10^{-6}$ and $\sigma_\alpha\simeq1250$, with the spectrum discretized
at 1000 equally spaced energies,
$\omega_i=10^{-4}+(i-1)(8-10^{-4})/1000$ for $i=1,\ldots,1000$.
In the correlator integral, the integration weights at the first and
last energy points are half the uniform spacing $\Delta\omega$.

\begin{figure}[!htbp]
 \centering
 \includegraphics[width=\columnwidth]{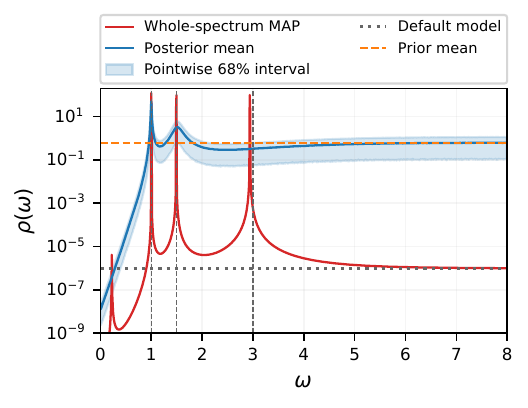}
 \caption{Three-peak mock reconstruction for $m=10^{-6}$, on a logarithmic
 scale. The red curve is the
 alpha-marginalized whole-spectrum MAP; the blue curve and band show
 the posterior mean and pointwise central 68\% interval. Vertical
 dashed lines mark the input delta-peak positions.
 The gray dotted line is the default model, and the orange dashed line
 shows the prior mean weighted with $p(\alpha\mid D)$, as given in
 Eq.~\eqref{eq:averaged_prior_mean}.}
 \label{fig:mock_spectrum}
\end{figure}

Figure~\ref{fig:mock_spectrum} compares the MAP, posterior mean
$\E[\rho_i\mid D]$ and pointwise central 68\% credible interval.
The MAP recovers three localized peaks near the input energies.
Their centroids are approximately 1.00, 1.49 and 2.94, and the
areas under the three peaks are 1.00, 0.98 and 0.98, respectively,
close to the input spectral weight of 1 for each state. The corresponding peak
areas in the posterior mean are 1.02, 0.89 and 0.33.
The reconstruction captures the main spectral structures. The MAP also
contains a small additional peak near
$\omega\simeq0.22$, with height approximately $4.20\times10^{-6}$,
where the input spectrum vanishes.

\begin{figure}[!htbp]
 \centering
 \includegraphics[width=\columnwidth]{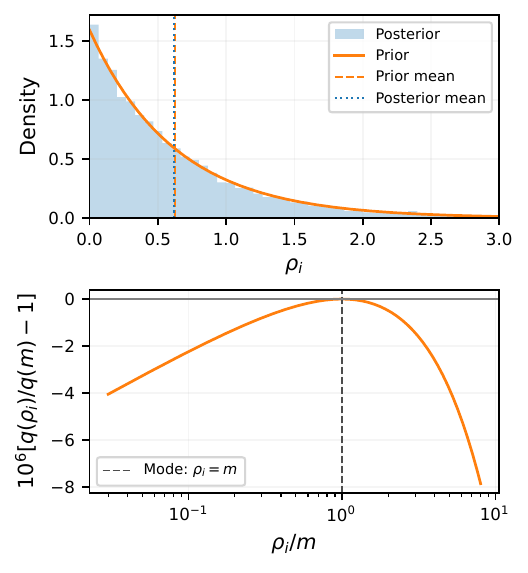}
 \caption{Prior and posterior densities for the baseline mock. Upper:
 the prior density $q(\rho_i)$, obtained by averaging over
 $p(\alpha\mid D)$, and the sampled high-energy posterior marginal at
 $\omega\simeq7.70$, with their means marked. Lower: to resolve the mode
 at $m=10^{-6}$, we plot $10^6[q(\rho_i)/q(m)-1]$ against $\rho_i/m$,
 magnifying the relative density variation by a factor of $10^6$.
 Histogram normalization includes samples outside the displayed range.}
 \label{fig:reference}
\end{figure}

The MAP selects the most probable spectrum, whereas the posterior mean
averages over the full posterior and is sensitive to its asymmetry and
extended tails. This distinction is closely related to the strongly
right-skewed BR prior. Each conditional prior has its mode at $m$, so
the prior averaged over $p(\alpha\mid D)$ also has its mode there.
Its mean follows from Eq.~\eqref{eq:prior_moments},
\begin{equation}
 \begin{aligned}
 &\int_0^\infty\dd\alpha\,p(\alpha\mid D)\E[\rho_i\mid\alpha]\\
 &\qquad=m_i\left[1+
 \frac{\E_{\alpha\mid D}[\alpha^{-1}]}{\Delta\omega_i}\right],
 \end{aligned}
 \label{eq:averaged_prior_mean}
\end{equation}
provided the inverse moment is finite. For the baseline mock, this mean
is approximately 0.63, as shown by the orange dashed line in
Fig.~\ref{fig:mock_spectrum}, while the mode remains at $m=10^{-6}$.
Figure~\ref{fig:reference} displays the right-skewed averaged prior
and its extended tail, with the mean marked in the upper panel and
the mode resolved in the lower panel.

In Fig.~\ref{fig:mock_spectrum}, the MAP and mean spectra tend toward
different baselines in the valleys between their respective peaks:
the prior mode $m$ and the weighted prior mean, respectively.
This distinction is particularly clear at high energies, where the
connection to the prior can be established directly. The exponential
kernel suppresses contributions from this region to the correlator.
Changes in the spectral values then have little effect on the
correlator within its statistical uncertainties, leaving these values
weakly constrained by the data. The posterior is therefore governed
by the prior, so its mean approaches the weighted prior mean and its
MAP approaches the prior mode $m$. In this regime, the posterior is
approximately given by the conditional prior averaged over
$p(\alpha\mid D)$,
\begin{equation}
 p(\rho_i\mid D)\simeq
 \int_0^\infty\dd\alpha\,p(\rho_i\mid\alpha)\,p(\alpha\mid D).
 \label{eq:weak_prior_limit}
\end{equation}
The upper panel of Fig.~\ref{fig:reference} illustrates this limit:
the high-energy posterior marginal at $\omega\simeq7.70$ closely
follows the averaged prior density.

These differences are consistent with previous discussions of posterior
sampling: skewness and tails can lead to different means, medians and
most probable values~\cite{Rothkopf2022}. The choice of a spectral
estimate depends on the properties one aims to recover~\cite{Burger2014}.
In this mock test, the MAP provides higher spectral resolution and
reproduces the input peak positions and areas more faithfully than the
posterior mean. Although the mean reflects additional aspects of the
posterior distribution, the MAP is
the better choice for representing the narrow spectral structures
studied here.

\subsection{Information-theoretic analysis of spectral constraints}

\begin{figure}[!htbp]
 \centering
 \includegraphics[width=\columnwidth]{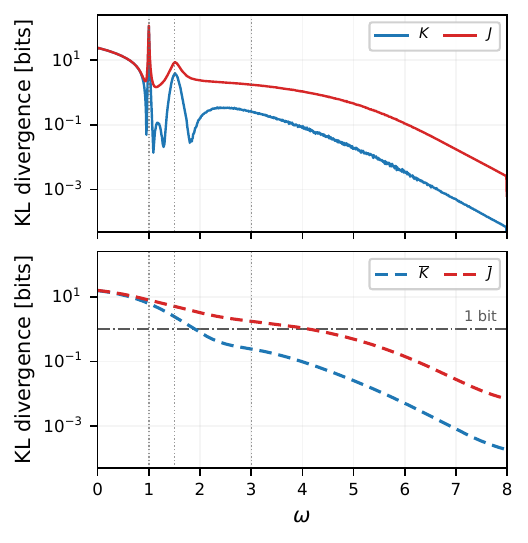}
 \caption{Information measures for the baseline mock. Upper:
 unsmoothed conditional KL $K$ (blue) and full-conditional KL $J$
 (red). Lower: the corresponding Gaussian-smoothed curves
 $\overline K$ and $\overline J$, with $\sigma=0.5$.
 The horizontal dash-dotted line marks a semiquantitative reference of 1 bit.
 Vertical dotted lines mark the three input peak positions.}
 \label{fig:mock_information}
\end{figure}

We now use $K$ and $J$ to quantify the data constraints underlying
Fig.~\ref{fig:mock_spectrum}. Their energy dependence is shown in
Fig.~\ref{fig:mock_information}. At the first peak, $\omega\simeq1.00$,
$K_i\simeq109.88$ bits and $J_i\simeq117.43$ bits: the data supply more than
100 bits of information about the spectral value relative to its
conditional prior. The correlation contribution $J_i-K_i$ in
Eq.~\eqref{eq:decomposition} is small compared with $K_i$, which
quantifies how much the data change the probability distribution of
$\rho_i$ from prior to posterior.
At the second peak, $\omega\simeq1.50$, $K_i\simeq3.61$ bits and
$J_i\simeq8.34$ bits. The data supply less information than at the
first peak, with approximately 4.73 bits encoded in correlations,
comparable to $K_i$. At the third peak,
$\omega\simeq3.00$, the values are $K_i\simeq0.25$ bits and
$J_i\simeq1.77$ bits. These values show a further decrease in the information
supplied by the data, with most of it now encoded in correlations with
the rest of the spectrum.
Below the first peak, $K_i$ ranges from approximately 10.72 to 24.08 bits
for $\omega<0.60$. These large values show that the data strongly
constrain the spectral strength to be small, illustrating that a region
without a peak can also carry substantial information relative to the prior.
Toward the upper end of the energy range,
$K$ and $J$ decrease to values of order $10^{-4}$ and $10^{-3}$ bits,
respectively, indicating little information beyond the conditional
prior. The posterior there closely follows the weighted prior, as shown
previously in Fig.~\ref{fig:reference}.

The local dips in $K$ reflect its definition as a difference between
posterior and prior distributions. On either side of a peak, the
reconstructed spectral strength changes between the small values in a
valley and the large values near the peak center. Along this transition,
the MAP or posterior mean approaches its prior counterpart. This suggests
that the posterior distribution may also locally approach its conditional
prior, providing a possible explanation for the dip in $K$. For example, as shown in
Fig.~\ref{fig:mock_information}, between the
valley to the left of the first peak and its center, a dip occurs at
$\omega\simeq0.95$, where $K_i\simeq0.05$ bits. Its proximity to the
crossing of the posterior and prior means in
Fig.~\ref{fig:mock_spectrum} is consistent with this interpretation.
At the same energy, $J_i\simeq2.57$ bits, showing that some constraints
encoded in correlations with the rest of the spectrum remain appreciable.
The correlation contribution makes $J$ smoother than $K$, although
shallow dips remain visible. Such dips reflect local agreement with
the prior and need not indicate a corresponding weakening of data
constraints across the surrounding energy region.

To average over these localized dips and obtain a regional view of the
data constraints, we smooth the information curves over an energy scale
$\sigma$,
\begin{equation}
 \overline X(\omega_i)=
 \frac{\sum_j \Delta\omega_j e^{-(\omega_i-\omega_j)^2/(2\sigma^2)}X_j}
 {\sum_j \Delta\omega_j e^{-(\omega_i-\omega_j)^2/(2\sigma^2)}},
 \quad X=K,J,
 \label{eq:smoothing}
\end{equation}
where $X_j=X(\omega_j)$. This averages the information values themselves,
rather than computing a divergence for an integrated spectral weight.
The lower panel of Fig.~\ref{fig:mock_information} shows the results
for $\sigma=0.5$: the narrow dips are averaged out, revealing the broad
decrease in marginal information with energy and the additional
constraints encoded in correlations. The smoothed $K$ and $J$ curves,
read alongside the reconstructed spectrum, provide a
semiquantitative basis for assessing reconstruction reliability across
energy regions. Using 1 bit as an illustrative reference, $\overline J$
suggests appreciable data constraints up to $\omega\simeq4$, consistent
with the reconstructed spectrum in Fig.~\ref{fig:mock_spectrum}.
These main spectral and information structures are also observed for
$m=10^{-8}$ and $10^{-4}$.

\section{Lattice-correlator results}
\label{sec:real}

We now apply the framework tested in the mock study to the spectral
reconstruction of a lattice proton correlator.

\subsection{Data and reconstruction setup}

We analyze a proton two-point correlator calculated on 81 gauge
configurations of the RBC/UKQCD $48\mathrm{I}$ ensemble. This
$N_f=2+1$ domain-wall-fermion ensemble
has volume $48^3\times96$, lattice spacing $a=0.1141(2)\,\mathrm{fm}$
and sea pion mass $m_{\pi,\mathrm{sea}}\simeq139\,\mathrm{MeV}$~\cite{Blum2016}.
We use overlap valence fermions~\cite{Neuberger1998} in a mixed-action
setup at the near-physical valence pion mass
$m_{\pi,\mathrm{val}}\simeq149\,\mathrm{MeV}$.
The correlator is computed with the Gaussian-smeared $Z_3$-noise grid
sources described in Ref.~\cite{Liu2021Scalar} and point sinks.
We reconstruct the spectrum from the ensemble mean and its estimated
full covariance matrix at Euclidean times $t/a=1,\ldots,15$.
The relative statistical errors are substantially larger than in the
mock study, especially at later times, reflecting the lower
signal-to-noise ratio near the physical pion mass. This allows us to
examine the framework with a less precise correlator.

\begin{figure}[!htbp]
 \centering
 \includegraphics[width=\columnwidth]{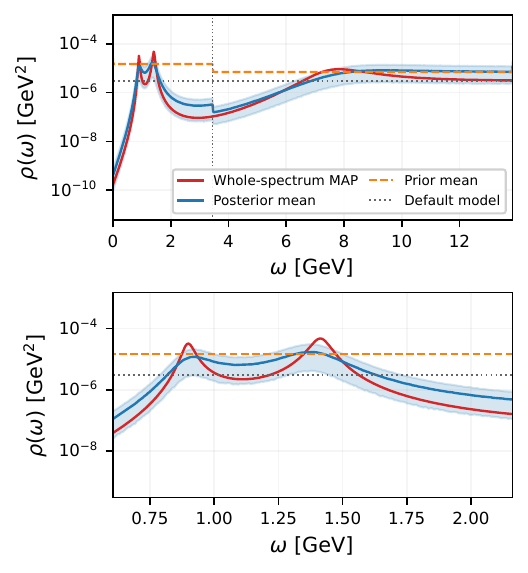}
 \caption{Proton spectral reconstruction on the $48\mathrm{I}$ ensemble. The red curve
 shows the whole-spectrum MAP; the blue curve and band show the
 posterior mean and pointwise central 68\% interval. The orange dashed
 curve is the prior mean weighted with $p(\alpha\mid D)$, and the gray
 dotted line is the default model. Upper: full energy range, with the
 vertical dotted line marking the change in energy spacing at
 $\omega\simeq3.46\,\mathrm{GeV}$. Lower: low-energy detail.}
 \label{fig:lattice_spectrum}
\end{figure}

We evaluate the spectrum at 1000 energies in $[10^{-4},8]\,a^{-1}$,
with spacings of approximately $0.0040\,a^{-1}$ below
$\omega=2a^{-1}$ and $0.0120\,a^{-1}$ above it, to improve resolution
in the low-energy peak regions.
We use the constant default model $m=10^{-6}a^{-2}$ and hyperprior scale
$\sigma_\alpha\simeq1250a$. Equation~\eqref{eq:averaged_prior_mean} gives
\[
 \E_{\alpha\mid D}\Bigl[\E[\rho_i\mid\alpha]\Bigr]-m
 =\frac{m\,\E_{\alpha\mid D}[\alpha^{-1}]}{\Delta\omega_i}.
\]
The roughly threefold increase in integration weights near
$\omega=2a^{-1}\simeq3.46\,\mathrm{GeV}$ therefore reduces the excess
of the averaged prior mean above $m$ by the same factor. The resulting
step in the orange dashed curve in Fig.~\ref{fig:lattice_spectrum}
is a discretization effect rather than a physical spectral structure.
From this perspective, the MAP may be a more appropriate choice for
displaying the reconstructed spectrum, since the prior mode $m$ is
unchanged by the integration weights.

Figure~\ref{fig:lattice_spectrum} compares the MAP and posterior mean
spectra. Both show two low-energy structures, with sharper peaks in the
MAP. The proton ground-state peak at $\omega\simeq0.90\,\mathrm{GeV}$
lies close to the physical mass of approximately $0.94\,\mathrm{GeV}$.
The second peak near
$\omega\simeq1.41\,\mathrm{GeV}$ may represent a combined contribution
from several unresolved nucleon excited states.
Above $\omega\simeq6\,\mathrm{GeV}$, no further spectral structures
are clearly resolved. Toward the upper end of the energy range, the
MAP and posterior mean approach
the prior mode $m$ and the averaged prior mean,
respectively. This high-energy behavior is consistent with the
weak-data limit discussed in Eq.~\eqref{eq:weak_prior_limit}.

\subsection{Information-theoretic analysis of spectral constraints}

\begin{figure}[!htbp]
 \centering
 \includegraphics[width=\columnwidth]{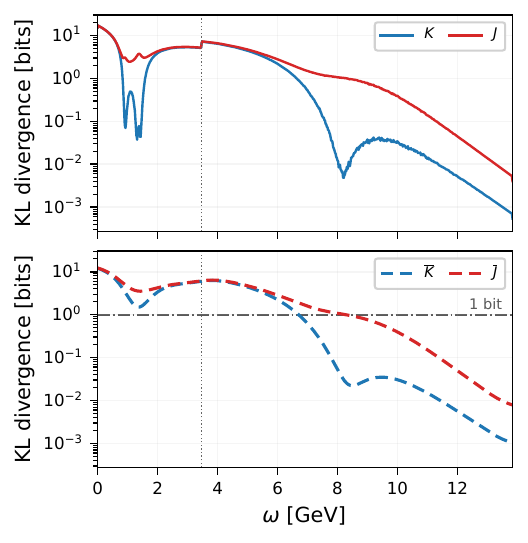}
 \caption{Information measures for the proton spectral reconstruction.
 Upper: unsmoothed conditional
 KL $K$ (blue) and full-conditional KL $J$ (red). Lower: the corresponding
 Gaussian-smoothed curves $\overline K$ and $\overline J$, with
 $\sigma=0.40\,\mathrm{GeV}$. The horizontal dash-dotted line marks the illustrative
 1-bit reference. Vertical dotted lines mark the change in energy
 spacing at $\omega\simeq3.46\,\mathrm{GeV}$.}
 \label{fig:lattice_information}
\end{figure}

We assess the data constraints on this reconstruction using $K$ and $J$,
shown in the upper panel of Fig.~\ref{fig:lattice_information}.
The lower panel shows their Gaussian averages at
$\sigma=0.40\,\mathrm{GeV}$, defined by Eq.~\eqref{eq:smoothing}.
Compared with the mock study, the information measures are overall
smaller and show less pronounced peak structures, consistent with
the reduced precision of the correlator data.

In the upper panel of Fig.~\ref{fig:lattice_information}, $K$ exhibits
pronounced local dips near the two low-energy peaks. At their centers,
$K_i\simeq0.09$ and 0.05 bits, indicating little change in the
individual posterior distributions from their conditional priors.
To understand the location of these dips, we turn to
Fig.~\ref{fig:lattice_spectrum}. In this lower-precision reconstruction,
the peaks are relatively low, and the posterior mean near their centers
remains close to the averaged prior mean. This is consistent with the
mechanism discussed in the mock study: local agreement with the prior
produces a dip in $K$, but here this agreement occurs near the peak
centers. Meanwhile, $J_i\simeq3.08$ and 3.67 bits at the same energies,
showing that constraints encoded in correlations with the rest of the
spectrum remain appreciable. To assess the constraints across each
peak region while averaging over these localized variations, we use
the Gaussian-smoothed measures in the lower panel. At the two peak energies,
$\overline K\simeq3.47$ and 1.51 bits, with
$\overline J\simeq5.07$ and 3.51 bits,
respectively. These values indicate that, across the peak regions,
the data appreciably modify the individual distributions relative to
their conditional priors, with additional constraints encoded in
correlations.

In the suppressed spectral region $2\lesssim\omega\lesssim6\,\mathrm{GeV}$,
$\overline K$ ranges from approximately 2.26 to 6.24 bits and
$\overline J$ from 2.87 to 6.38 bits. The two measures remain comparable
throughout this region. The data strongly constrain the spectral
strength to be small, with most of the information already expressed
in the distribution at each energy.

In the high-energy region $\omega>6\,\mathrm{GeV}$,
both smoothed measures decrease overall: $\overline K$ falls from a few
bits to $O(10^{-3})$ bits, while $\overline J$ falls from a few bits to
$O(10^{-2})$ bits toward the upper end of the reconstructed energy
range. This indicates progressively weaker data constraints at higher
energies, consistent with the mock study. Using the illustrative
1-bit reference, $\overline J$ falls
below this level near $\omega\simeq8.25\,\mathrm{GeV}$ in the lower
panel of Fig.~\ref{fig:lattice_information}. This gives a
semiquantitative estimate of the energy range appreciably constrained
by the data, with limited information beyond the prior above roughly
$8\,\mathrm{GeV}$, in good agreement with the approach of the
reconstructed spectra to their prior baselines in
Fig.~\ref{fig:lattice_spectrum}.

\section{Summary and outlook}
\label{sec:summary}

We have developed an information-theoretic framework for analyzing
spectral reconstruction from lattice correlators. The conditional KL
$K$ quantifies the information supplied by the data through the change
from the conditional prior to the posterior distribution of each
spectral value. The full-conditional KL $J$ measures this information
while also accounting for correlations with the rest of the spectrum.
Gaussian smoothing provides a regional assessment at a specified
energy scale.

We have evaluated these measures in both a three-peak mock study and a
lattice proton correlator calculation. The results demonstrate that
these divergences help explain the properties of reconstructed spectra
from an information-theoretic perspective. They also provide a
semiquantitative criterion for assessing how strongly the reconstructed
spectral structures are constrained by the data and how reliable they
are in light of the prior's influence.

In general, information-theoretic approaches offer a promising direction
for analyzing lattice correlators and reconstructing spectral functions.
They can help clarify the physical information encoded in spectral
functions and deepen our understanding of the inverse problem itself,
guiding the development of improved reconstruction algorithms.
Building on the present study, more sophisticated information-theoretic
formulations could provide a more detailed and accurate characterization
of spectral reconstructions. Our planned work will focus on using these
developments to guide prior selection, quantify systematic uncertainties
associated with prior choices, and establish more precise
information-theoretic criteria for reconstruction reliability.

\section*{Acknowledgments}
We thank Alexander Rothkopf for generously privde us his code of BR reconstruction and for instructive discussion.
The numerical calculations are performed on the Southern Nuclear Science Computing Center (SNSC) of South China Normal University and the Xiangjiang-1 cluster at Hunan Normal University (Changsha).
This work is supported by the National Natural Science Foundation of China (NSFC) under Grants No.\ 12575085, No.\ 12635009, No.\ 12675103 and No.\ 12547185.

\bibliography{references}

\end{document}